\documentclass{aa}
\usepackage{epsfig}

\newcommand{\bmv}{B$-$V}
\newcommand{\bmo}{(B$-$V)$_0$}
\newcommand{\teff}{T$_{\rm eff}$}
\newcommand{\nli}{$\log$~n(Li)}

\begin{document}
\title{Evolution of lithium beyond the solar age: a Li survey of
the old open cluster NGC~188
\thanks{Based on observations collected at the Italian National Telescope
Galileo}}

   \subtitle{ }

   \author{S. Randich\inst{1} \and P. Sestito\inst{2} 
         \and R. Pallavicini\inst{3}}

   \offprints{S. Randich, email:randich@arcetri.astro.it}

\institute{INAF/Osservatorio Astrofisico di Arcetri, Largo E. Fermi 5,
             I-50125 Firenze, Italy
\and
Dipartimento di Astronomia, Universit\`a di Firenze, Largo E. Fermi 5,
            I-50125 Firenze, Italy
\and
INAF/Osservatorio Astronomico di Palermo, Piazza del
                  Parlamento 1, I-90134 Palermo, Italy}

\titlerunning{Li in NGC~188}
\date{Received Date: Accepted Date}

\abstract{We have determined Li abundances for 11 G--type stars in
the 6--8~Gyr old open cluster \object{NGC~188}. 
These data significantly enlarge the number of cluster stars with
Li measurements,
allowing us to extend the investigation of Li depletion in
open clusters to ages well beyond the age of the Sun. We have 
also inferred the cluster metallicity which turns out to be solar.
We find that
solar--type stars in NGC~188 are only slightly more Li depleted
than the much younger \object{Hyades} and no more Li 
depleted than stars of similar
temperature in the 2--4~Gyr younger cluster \object{M~67}.
At variance with M~67, NGC~188 members show virtually no scatter in
their Li abundances. Surprisingly,  no solar--type star in NGC~188 appears as
Li depleted as the Sun or as the most Li depleted stars in M~67. We discuss
the implications of these results for mechanisms of internal
mixing and Li depletion in main sequence stars. 
}
\maketitle
\section{Introduction}
Observations of light elements in stars are powerful tools to test
Big Bang nucleosynthesis, Galactic chemical evolution and stellar
interior mixing processes.
Lithium has a very low burning temperature ($\mathrm{\sim 2.5\times 10^{6}\,K}$)
and it survives only in the outermost layers of a star;
therefore, it is a good tracer of mixing mechanisms and stellar 
interior structure.
Observations of Li in clusters of different ages and chemical
compositions allow us to put constraints on these processes
and to investigate their dependence on age, metallicity and possibly other
stellar parameters.
Observations of cluster stars carried out during the past 15 years 
have shown that
Li depletion is not a simple function of age and metallicity as predicted
by standard models (see e.g. the reviews of Jeffries \cite{jef00} and Pasquini
\cite{pas00} and references therein). For example, standard models,
which include only convection as a mixing process, predict that little (if any)
Li depletion should occur during the main sequence (MS) evolution of
solar--type stars, since the convective
zone is too shallow to reach the Li burning layer; the comparison of
the Li abundance distributions of clusters of different 
ages (spanning the range
from the zero age main sequence -ZAMS- to the solar age) 
indicates instead that solar--type stars do destroy Li on the MS. 
\begin{figure}
\psfig{figure=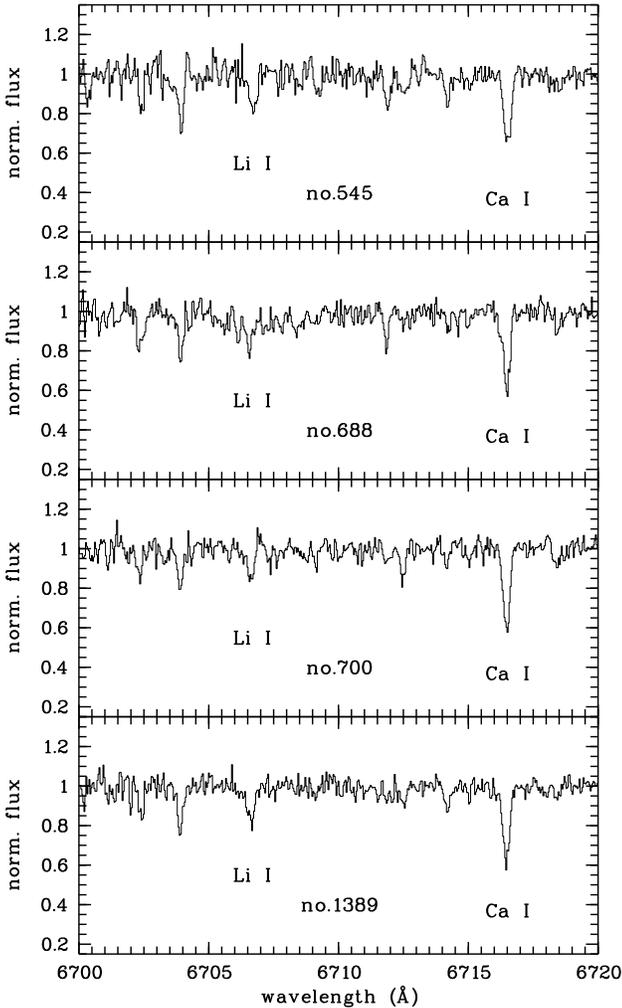, width=11.0cm, angle=0}
\vspace{-2cm}
\caption{Sample spectra in the Li region.}
\end{figure}
Furthermore,
the solar age, solar-metallicity cluster M~67 is characterized by
a large dispersion in Li abundances for stars
of similar temperatures (Spite et al. \cite{spi87}; Garc\'\i a L\'opez et al.
\cite{gar88}; Pasquini et al. \cite{pas97}; Jones et al. \cite {jon99}); 
about 60 \% of the stars appear Li-rich, with a Li abundance a factor of 2--3
below the 600~Myr old Hyades, 
while the remaining fraction are Li--poor with
a factor of 5--10 lower abundances. This result shows that
additional parameters besides mass, age, and metallicity
affect Li depletion, and that 
non-standard mixing processes such as diffusion, mass loss, or rotationally
driven mixing, are at work.

The comparison between M~67 and the intermediate age clusters 
\object{IC~4651},
\object{NGC~3680}, and \object{NGC~752} (age $\sim 2$~Gyr) 
shows instead that stars in the upper envelope of M~67 have similar abundances
as the intermediate cluster stars,
suggesting that either Li depletion becomes very slow at old ages, 
at least for a fraction of stars, or that
different conditions (e.g., heavy element
abundances, initial angular momentum distributions and/or rotational
evolutions) in different clusters lead to different Li evolutions
(Randich et al. \cite{r00}). We also mention that neither IC~4651 nor
NGC~3680 show any scatter in Li, while a scatter might be present
among NGC~752 members cooler than 5800~K (Hobbs \& Pilachowski \cite{hp86}). 

In order to put additional empirical constraints
on MS Li depletion, in particular on its timescale and
on the development of the dispersion, we carried out
a Li survey of the old open cluster NGC~188. This cluster 
has an estimated
age of 6--8~Gyr (see discussion in Sarajedini et al. \cite{sara99},
hereafter S99) and is one of the oldest known open clusters.
According to S99 it is $3.0\pm 0.7$~Gyr older than M~67.
The only
published data on Li abundances in dwarf stars in this cluster
date back to nearly 15 yrs ago and were limited to only a few
stars (Hobbs \& Pilachowski \cite{hp88}).
A more detailed Li study of this cluster is thus warranted.
Our paper is organized as follows: in Sect.~2 we describe our sample
and the observations; the abundance analysis and the results are summarized 
in Sects.~3 and 4, while in Sects. 5 and 6 we present a discussion
of the results and our conclusions.
\section{Observations}
We selected the target stars from the proper motion study of Dinescu et al.
(\cite{din96} --hereafter D96); 
our sample includes 11 G--type stars with membership probability
larger than 90\%. The sample stars are listed in Table~1; in the first and
second columns the numbering systems of D96 and S99 are used.
In the following we will use the numbering system of D96.
Three stars are in common with the study of Hobbs \& Pilachowski (\cite{hp88}).
The observations were carried out in service mode during August 2001
at the Italian National Telescope Galileo
(TNG) equipped with the SARG spectrograph (Gratton et al. \cite{grat00}).
The Red Grism and
the OG570 filter were used, together with the mosaic of two EEV 15~$\mu$
pixel CCDs
(2048 $\times$ 4096); such a combination allowed us to cover
the spectral range from $\sim 550$
to 1011~nm. 2$\times 2$ CCD binning was used.
Part of the sample stars were observed using a slit width
equal to 0.8$^{"}$, corresponding to a nominal resolving power R$\sim 
57,000$, while another fraction was observed with a larger slit
(1.6$^{"}$). In the latter case the seeing was significantly better than the
slit width and, therefore, we obtained a resolving power 
R$ \sim 35,000$ (as measured from the telluric lines), somewhat higher
than the nominal one (R$\sim 29,000$).

The spectra were reduced using MIDAS in the ECHELLE context.
We first separated, rotated, and flipped the two CCD; then, we carried
out the reduction following the usual steps: namely, bias subtraction,
order definition,  order extraction,
inter-order background subtraction, flat-fielding and wavelength calibration.

All the sample stars have V magnitudes around 15 and all but two of
them (no.~1344 and no.~700) 
were exposed for 2 hrs; exposure times
for stars no.~1344 and 700 were slightly shorter (1.5 and 1.75 hrs, 
respectively). Typical S/N
ratios in the lithium region are in the range 20--35.
The spectra of four stars in the sample are shown in Fig.~1.
%
\begin{table*}[!ht] \footnotesize
\caption{Stellar parameters and Li abundances for the sample stars. Asterisks
indicate objects with nominal resolving power R$=57,000$.} 
\begin{tabular}{cccccccc}
\hline
\hline
no. & no. & V & B--V & \teff & $\mathrm{EW({\lambda}~6708~\AA)}$ & \nli$_{\rm LTE}$ & \nli$_{N\rm LTE}$\\
D96 & S99 & & & $\mathrm{(K)}$ & $(\mathrm{m\AA})$ & & \\
\hline
545$^*$  & 989 & 14.99 & 0.74 & 5696 & 60$\pm$7   &   2.29    & 2.30$\pm$0.11\\
612$^*$  & 1172 & 15.05 & 0.69 & 5888 & 34$\pm$6   &   2.19    & 2.19$\pm$0.11\\
661$^*$ & 1025  & 15.10 & 0.67 & 5967 & 64$\pm$7   &   2.57    & 2.54$\pm$0.11\\
688 &724  & 15.13 & 0.67 & 5967 & 46$\pm$7   &  2.34     & 2.38$\pm$0.10\\
700 & 685 & 14.97 & 0.69 & 5888 & 40$\pm$6 & 2.26      & 2.26$\pm$0.10\\
760 &534  & 15.16 & 0.66 & 6008 & 42$\pm$6   &  2.39     & 2.37$\pm$0.11\\
1344 & 1120 &  14.92 & 0.72 & 5772 & 31$\pm$6   &   2.05    & 2.06$\pm$0.12\\
1354$^*$ & 933 &  15.05 & 0.68 & 5928 & 69$\pm$8   &  2.58     & 2.55$\pm$0.11\\
1389&739  &  14.97 & 0.71 & 5810 & 53$\pm$7   &   2.33    & 2.33$\pm$0.11\\
1403$^*$ & 356 &  15.06 & 0.68 & 5928 & 39$\pm$6   &  2.29     & 2.28$\pm$0.11\\
1422$^*$ & 922 &  15.10 & 0.68 & 5928 & 49$\pm$9 & 2.40      & 2.38$\pm$0.13\\

 \hline
\end{tabular}
\end{table*}

\begin{table*}[!ht] \footnotesize
\caption{Stellar parameters and Li abundances for the Hobbs \& Pilachowski
(\cite{hp88}) sample.}
\begin{tabular}{ccccccccc}
\hline
\hline
no. & no. & no. & V & B--V & \teff  & $\mathrm{EW({\lambda}~6708~\AA)}$ & \nli$_{\rm LTE}$ & \nli$_{\rm NLTE}$\\
D96 &San62 & S99 & & & $\mathrm{(K)}$ & $(\mathrm{m\AA})$ & & \\
\hline
661 &I-71&1025 & 15.10 & 0.67 & 5967 & 40$\pm$15 & 2.34 & 2.32$\pm$0.20\\
688 &I-101&724 & 15.13 & 0.67 & 5967 & 40$\pm$15 & 2.34 & 2.32$\pm$0.20\\
694 &677 & I-99  & 14.97 & 0.72 & 5772 & 72$\pm$15 & 2.46 & 2.46$\pm$0.15\\
700 &I-91&685  & 14.97 & 0.69 & 5888 & 25$\pm$15 & 2.05 & 2.06$\pm$0.33\\
726 &I-17& 735 &  15.11 & 0.67 & 5967 & 40$\pm$15 & 2.34 & 2.32$\pm$0.20\\
 \hline
\end{tabular}
\end{table*}

\begin{table*}[!ht] \footnotesize
\caption{Fe~{\sc i} equivalent widths and iron abundances from the two spectra
obtained by adding the observed spectra of stars with similar colors. The
stars used for the co-added spectra are indicated.}
\begin{tabular}{cccccc}
\hline
\hline
& \multicolumn{2}{c}{1354+1403+1422} & \multicolumn{2}{c}{612+700} \\
$\lambda$ (\AA) & EW & $\log{\mathrm{\epsilon(Fe)}}$ & EW & $\log{\mathrm{\epsilon(Fe)}}$\\
& (m\AA) & & (m\AA) & \\
& & & & \\
6703.58 &34 &7.59 & 33 & 7.53\\
6710.32 &12 &7.51 & 10 & 7.38 \\
6726.67 &42 &7.48 & 41 & 7.44 \\
6733.15 &28 &7.59 & 31 & 7.64 \\
6750.16 &70 &7.54 & 74 & 7.54 \\
6806.86 &27 &7.44 & 33 & 7.54 \\
6810.27 &46 &7.50 & 51 &7.57 \\
6820.37 &45 &7.64 & 42 &7.56 \\
6843.65 &--- &--- & 60 &7.53 \\
average &  & 7.53$\pm 0.07 \pm 0.09$ & & 7.53 $\pm 0.08 \pm 0.09$\\ 
\hline
\end{tabular}
\end{table*}
\section{Abundance analysis}
With the exception of star no.~545, B--V and V values for all our
sample stars were taken from
the UBV CCD photometric study of S99; 
actually, since in that paper photometry for each star is not listed,
we retrieved the values from the WEBDA
database\footnote{http://obswww.unige.ch/webda/}.
For star no.~545 we adopted the photographic
photometry of D96
(B--V=0.74, V=14.99), since we found
significantly discrepant values in S99 (B--V=0.45, V=19.50):
D96 photometry is most likely the correct one, based on the 
spectral features and Li content of this star,
which are similar to those of the other observed objects. 
Moreover,
we would have not been able to observe an object much fainter 
than $\mathrm{V\sim15}$. 
Photometry for the sample stars is listed in Cols. 3 and 4 of Table~1.

Effective temperatures were determined from B--V colors using 
the calibration of Soderblom et al. 
(\cite{sod93}). A reddening E(B--V)$=0.09$ was used (S99).
A surface gravity $\log g=4.44$ was assumed for all sample
stars, while microturbulence values were computed as
$\xi=3.2 \times 10^{-4} (\rm T_{\rm eff} -6390)-1.3(\log g-4.16)+1.7$
(Boesgaard \& Friel  \cite{bf90}). We recall that gravity and microturbulence
values have very little effects on the determination of Li abundances, while
they affect iron abundances. We assumed conservative random errors of
100~K in \teff, 0.3~dex in $\log g$, and 0.3~km/s in $\xi$.

Li abundances were derived from measured
equivalent widths of the
Li I $\mathrm{{\lambda}6707.8\, \AA}$ blend and by using the curves of growth
(COG) of Soderblom et al. (\cite{sod93}).
Li abundances were then corrected for NLTE effects using the prescriptions
of Carlsson et al. (\cite{carl94}). 
We mention that at our spectral resolutions and 
given the slow rotation of the sample stars, we are able to separate
the Li feature from the nearby  Fe~{\sc i}~$\mathrm{{\lambda}6707.44\, \AA}$,

Starting from published EWs and colors, we re--analyzed
in a consistent way the NGC~188 stars observed by Hobbs \& Pilachowski
(\cite{hp88}). Photometric
data for these stars, which are listed in Table~2 using three numbering
systems (D96, San62 --Sandage \cite{san62} and S99),
were retrieved from S99. 
Since we want to compare the \nli~vs. \teff~distribution of NGC~188 with
both the Hyades and M~67, abundances for these two clusters were also
re-determined consistently with our sample stars. Namely, for the Hyades, 
using colors and equivalent widths published by
Balachandran (\cite{bala95}), we derived new \teff~and Li abundances.
Jones et al. (\cite{jon99}) had derived Li abundances for M~67 using
the same temperature vs. color calibration and the same COGs we used in 
the present analysis and thus we just
applied NLTE corrections to their published abundances.
Errors in Li abundances for our sample stars were determined considering
both errors in EWs and uncertainties in effective temperatures.
Measured EWs, effective temperatures, and derived Li abundances are
listed in the last three columns of Table~1. The same quantities for
the sample of Hobbs \& Pilachowski are listed in Table~2.

The S/N ratios of our spectra are too low to allow a very accurate
determination of
the cluster metallicity. In order to have at least a rough estimate
of the cluster [Fe/H] from our spectra, and in particular 
to investigate whether or not it is
significantly higher/lower than solar,
we added the spectra
of stars with the same colors and measured the equivalent widths
of various Fe~{\sc i} lines in the same echelle order as lithium
and in the next (towards the red) order.
More specifically, we added the spectra of stars no.~1354, 1403, and
1422 (B--V$=0.68$, \teff$=5928$~K) and those of stars no.~612 and 700
(B--V$=0.69$, \teff$=5888$~K).
We then measured the EWs of 8-9 Fe~{\sc i} lines, listed in Table~3,
in the co-added spectra.
Iron abundances were determined using MOOG (Sneden \cite{sn73} 
--version December 2000) and Kurucz (\cite{kur95})
model atmospheres. Since we wanted to carry out
a differential analysis with respect to the Sun, we determined $\log gf$ 
values by an inverse abundance analysis of
the solar spectrum (sky at twilight) obtained with SARG during
our run, under the condition $\log \epsilon(\rm Fe)_{\odot}=7.52$; the
following solar parameters were assumed: \teff=5770~K, $\log g=4.44$, and
$\xi=1.1$~km/sec.

In Table~3 we list iron abundances for the different lines, together with
the mean value for each group of stars. The error includes the standard
deviation in the mean from the different lines (which provides an estimate
of the random error due to errors in the measured equivalent widths) 
and the random error due
to the uncertainties in stellar parameters. The latter was estimated by varying
one parameter at the time and leaving the other two parameters unchanged;
the errors due to each parameter were then quadratically added.
Uncertainties 
of $\pm 100$~K in \teff, $\pm 0.3$~dex in $\log g$, and $\pm 0.3$~km/s,
correspond to 0.07, 0.05, and 0.02 dex in $\log \epsilon$(Fe).
We find the same
iron abundance for the two group of stars; the weighted mean is $\log
\epsilon (\rm Fe) = 7.53 \pm 0.08$, i.e., NGC~188 has solar metallicity.
Whereas systematic errors due e.g., to reddening, effective temperature
scale, model atmospheres, atomic parameters, etc. are difficult to
estimate, we note that
our result is in agreement with earlier metallicity estimates based on
spectroscopy (of warner stars at the turn--off)
or photometry (see discussion in von Hippel \& Sarajedini
\cite{von98}).
\section{Results}
\subsection{Li equivalent widths}
\begin{figure*}
\psfig{figure=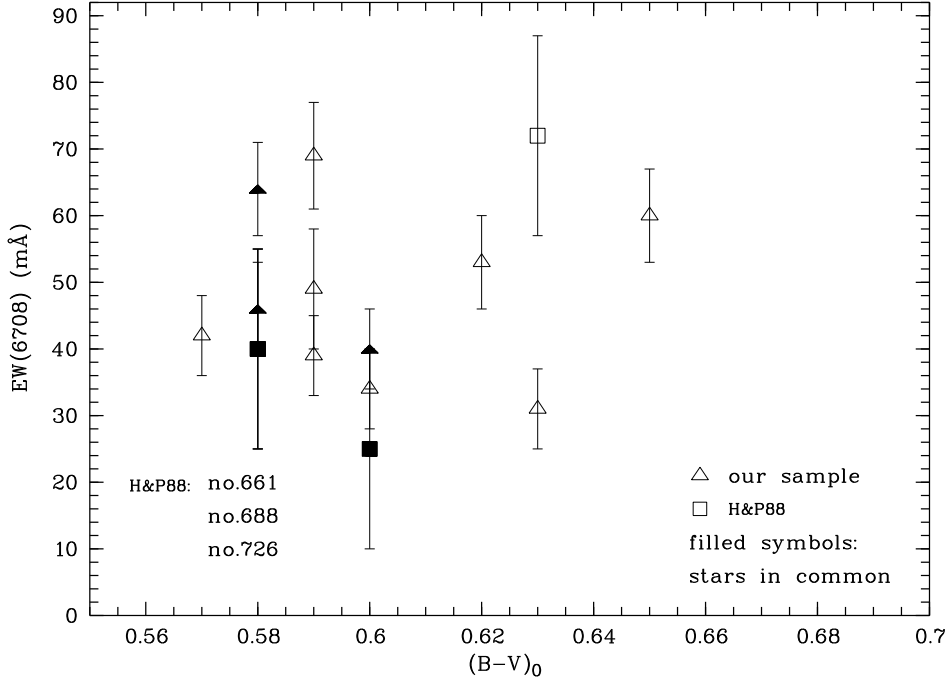, width=14.0cm, angle=-90}
\caption{Li EWs vs. dereddened B--V color for our sample stars
(triangles) and the stars in the sample of Hobbs \& Pilachowski
(\cite{hp88} --squares). Filled symbols denote stars in common
between the two samples. Note that, as indicated in Table~2,
stars no.~661, 688, and 726 in Hobbs \& Pilachowski sample
have the same color and Li EW. Of these, star no.~726 is not included
in our sample.}
\end{figure*}
In Fig.~2 we plot Li EWs as a function of dereddened \bmv~color for our sample
stars (triangles) and for the sample of Hobbs \& Pilachowski 
(\cite{hp88} --squares).
Filled symbols denote the three stars in common between the two samples.
The figure shows a good agreement between the EW vs. color distribution of
our sample stars and that from Hobbs \& Pilachowski; the EWs of the stars
in common are consistent within the errors, although our EWs are somewhat
larger than Hobbs \& Pilachowski's. In the following analysis, therefore,
we will merge the two samples, using our own Li measurements for the stars
in common. The total sample contains 13 stars, 11 stars observed by us plus
the two stars from Hobbs \& Pilachowski not included in our sample.
Figure~2 also evidences a certain amount
of star-to-star scatter. However the spread is consistent with the
observational errors: the average EW of our 11 measurements and of the two
stars from Hobbs \& Pilachowski that are not in our
sample is 49~m\AA~ with
a standard deviation $\sigma=13$~m\AA, that is only slightly larger
than the average EW error (8~m\AA). In addition, as we will see in the next
section, the
spread in EWs translates into a negligible dispersion in Li abundances.
\subsection{Li abundance vs. \teff}
\begin{figure*}
\psfig{figure=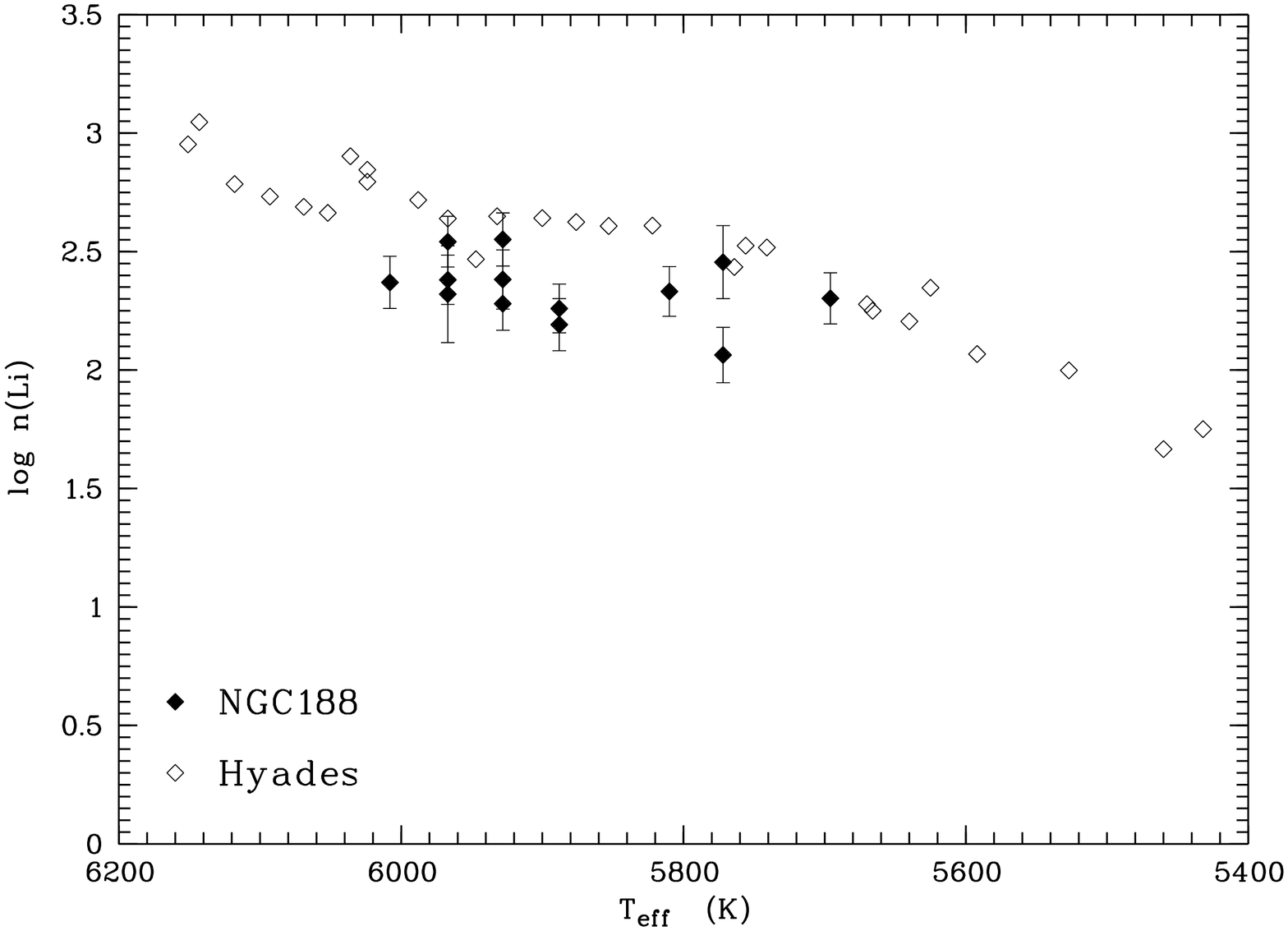, width=14.0cm, angle=-90}
\caption{Li abundances vs. \teff~for the 13 stars in the merged (our
$+$ Hobbs \& Pilachowski's) NGC~188 sample (filled symbols) and the Hyades
(open symbols).}
\end{figure*}
\begin{figure*}
\psfig{figure=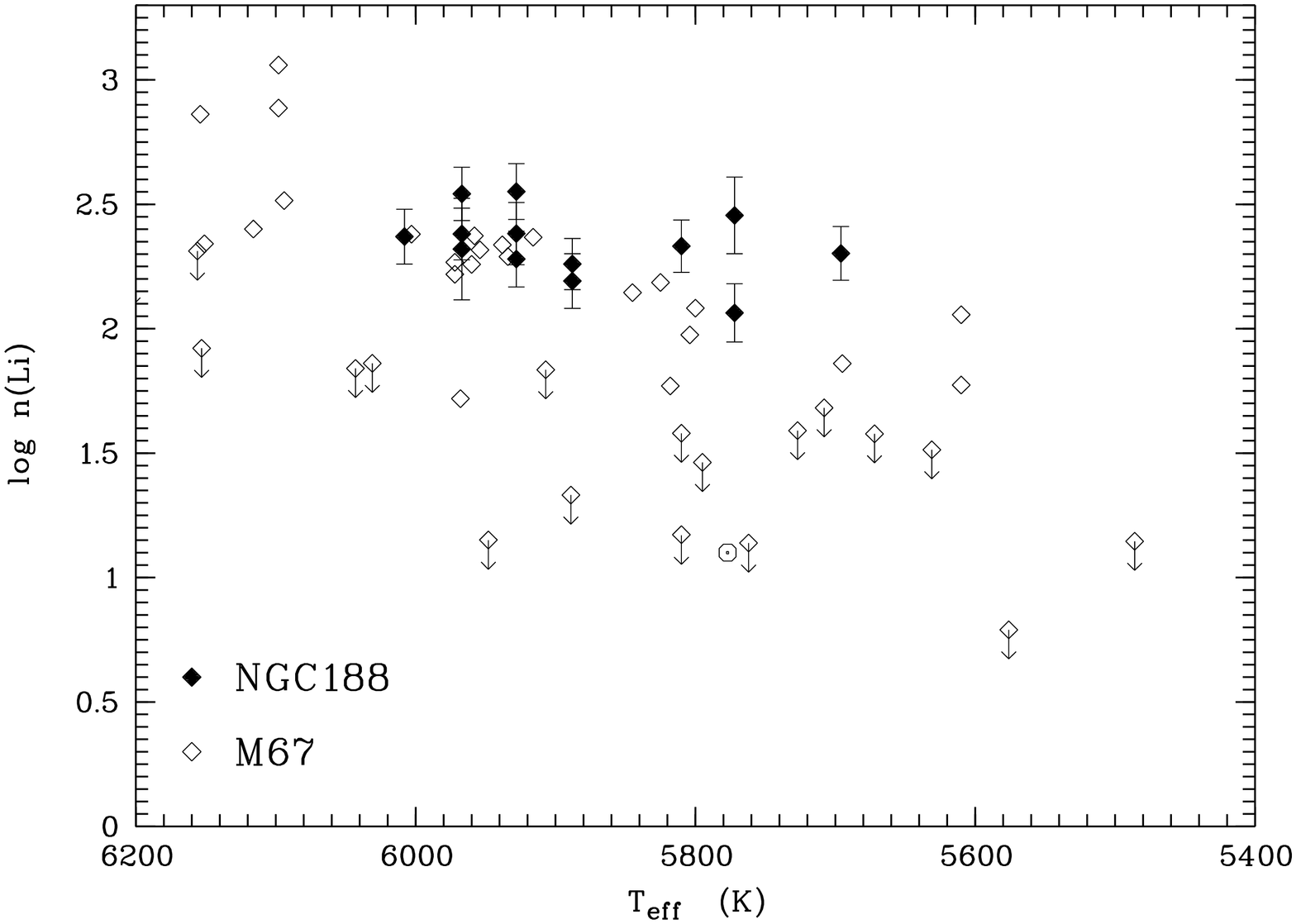, width=14.0cm, angle=-90}
\caption{Same as Fig.~3 but with NGC~188 compared to M~67. The Sun is also
shown in the figure.}
\end{figure*}
The usual \nli~vs. \teff~plot is shown in Figs.~3
and 4; in the two figures NGC~188 is compared to the Hyades and M~67,
respectively. As we have mentioned in Sect.~3, data for both clusters
were retrieved from the literature and analyzed consistently with
our sample stars.

The distribution of NGC~188 lies somewhat below the Hyades, but not
as much as one would expect given the difference in age
(a factor of at least 10). In particular, we note a couple 
of stars in our sample
with abundances similar to those of Hyades stars with the same
effective temperature. For the 13 NGC~188 stars we obtain
an average Li abundance
\nli$=2.34 \pm 0.13$; for the Hyades in the same temperature range
we obtain \nli$=2.58 \pm 0.09$, implying a difference in Li depletion
less than a factor of two.

The comparison with M~67 shows that:
{\it a)} the \nli~vs. \teff~distribution for NGC~188
lies on the upper envelope
of the pattern of M~67, i.e., 6--8~Gyr old members of NGC~188
are no more Li depleted than the Li-rich stars in the
younger M~67; viceversa,
several M~67 members are more Li depleted than NGC~188 stars. Whereas
this result was already pointed out by Hobbs \& Pilachowski (\cite{hp88}),
our observations allow us to confirm it based on a much larger sample
of stars;
{\it b)} NGC~188 does not show a
star-to-star scatter as large as that observed among M~67 members;
whereas the coolest stars in NGC~188 may be characterized by some
amount of scatter, the scatter is much smaller than the dispersion
of M~67 and most likely due to observational uncertainties.
Our NGC~188 sample is much smaller than the M~67 sample and
covers a narrower temperature interval; thus, the obvious
question arises whether the lack of a significant dispersion is simply
due to low number statistics and/or to the restricted temperature (mass)
range. In order to address this issue, we carried
out a simple statistical test; given the fraction of Li-poor 
stars in M~67 (we assumed a conservative value of 30 \%, although
according to Pasquini et al. \cite{pas97} it may be as high as 40 \%), 
and assuming that the same fraction would hold also for NGC~188,
we computed the probability of finding 0/13 Li-poor
stars in NGC~188. 
Using a binomial distribution,
we found {\bf P}(0/13)$\sim 9.7\times 10^{-3}$; this means
that we can exclude 
the possibility of the presence of Li-poor stars (and thus of a scatter)
in NGC~188 at almost the 3$\sigma$ significance level. Considering a fraction
Li-poor stars equal to 40 \%, 
we would get a probability {\bf P}(0/13)$\sim 1
\times 10^{-3}$ (larger than 3$\sigma$). In addition, even
considering only stars in the temperature range covered by our NGC~188 sample,
a scatter is present among M~67 members, with about
40~\% of the stars being Li-poor (see Fig.~4). 
Thus we can safely exclude that the lack
of a scatter in NGC~188 is due to the narrow temperature range.
\section{Discussion}
As a summary of the results presented in
the two previous sections we conclude that: 1) G--type stars in
NGC~188 are no more Li depleted than Li-rich stars in M~67 and are 
at most a factor of two more Li depleted than their Hyades counterparts;
2) M~67 remains so far the only cluster showing
a clear spread in Li among solar--type stars \footnote{As mentioned in the 
introduction, NGC~752 might show a spread among stars cooler than the
Sun (Hobbs \& Pilachowski \cite{hp86}), but this result
is based on two cluster members only and must be confirmed with
a larger sample.}.
In other words, the evolution of the upper envelope of the Li vs. \teff~
distribution past the age of the Hyades, as well as the evolution of the spread 
from the age of M~67 to that of NGC~188, further challenge
our understanding of the Li destruction in solar--type stars.
\subsection{The upper envelope}
\begin{figure*}
\psfig{figure=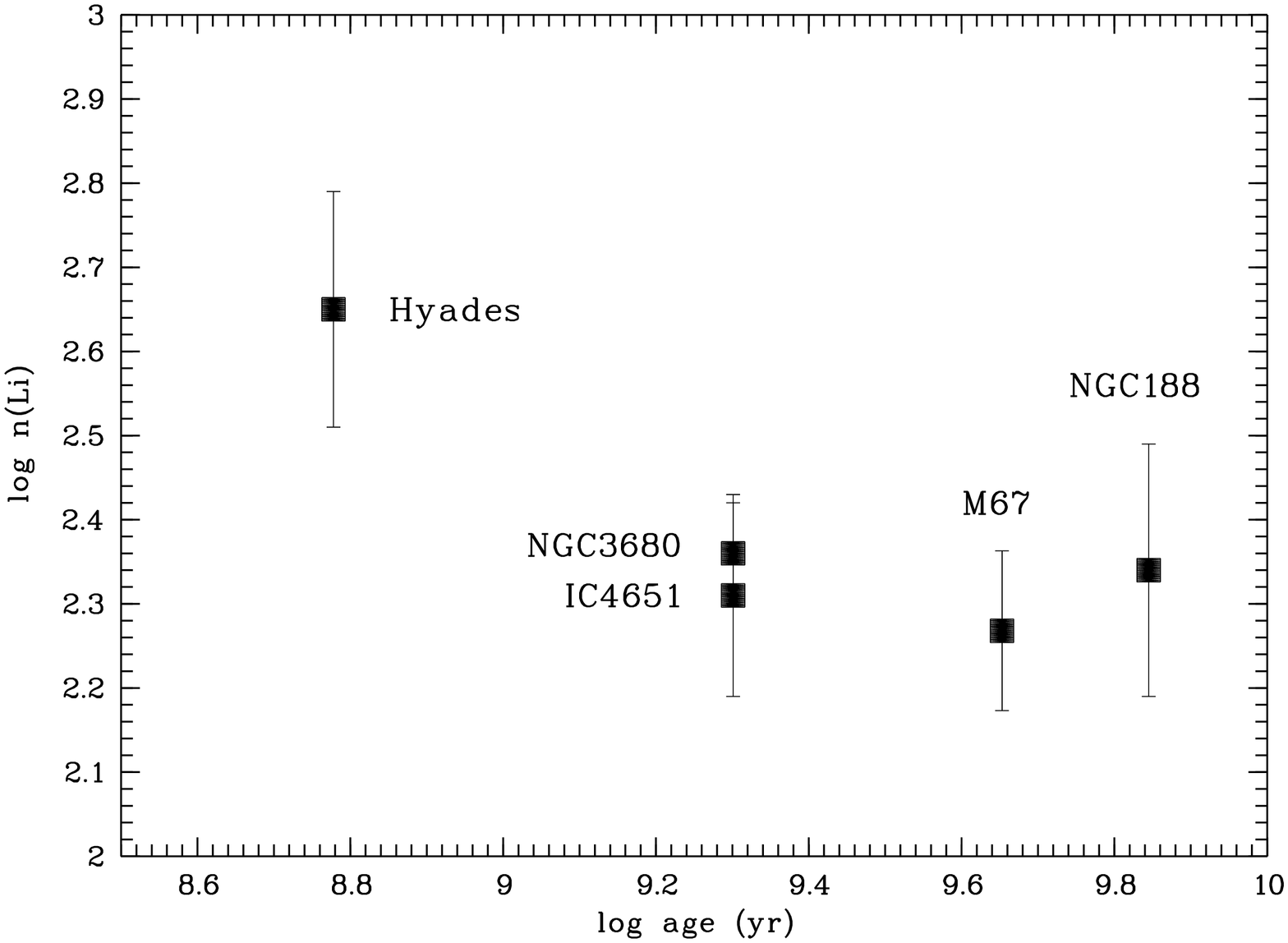, width=14.0cm, angle=-90}
\caption{Average Li abundance vs. age for five open clusters.
The average \nli~values have been computed considering
stars in the temperature range
$\mathrm{5750\leq{T_{eff}}\leq6050}$ K. Error bars
denote $1\sigma$ standard deviations. NGC~752 is not included
in the plot since only two stars in the considered \teff~range have
available Li measurements. Note however that for these two stars
we derive Li abundances $\log$~n(Li)$=$ 2.32 and 2.43 (as for the
other clusters, we have re-analyzed the original data of
Hobbs \& Pilachowski \cite{hp86}).}
\end{figure*}
As we mentioned in the introduction, Randich et al. (\cite{r00}), 
on the basis of a comparison
of the Hyades, of the intermediate age clusters IC~4651, NGC~3680,
and NGC~752, and of the solar age cluster M~67, concluded that
the mechanism that drives Li depletion appears to stop 
in the late phases ($\geq$~2~Gyr) of MS evolution, unless different
conditions/parameters lead to different Li depletion timescales.
In Fig.~5 we plot the average Li
abundance as a function of age for stars in the 5750--6050~K
temperature range. The Hyades, IC~4651 and NGC~3680, M~67, and NGC~188
are considered in the plot; for M~67 we included only stars lying
on the upper envelope (\nli~$\ge 2$). Error bars 
represent $1 \sigma$ standard deviations from the average.
The figure evidences a plateau
in Li abundances for ages older than $\sim 2$~Gyr and, except for Li--poor
stars in M~67, there seems to be no additional evolution of Li abundances
beyond 2~Gyr.
This result suggests three alternative scenarios:
{\it i)} the clusters have all the same initial abundance, but
solar--type stars have not undergone additional Li depletion beyond
$\approx 2$~Gyr, i.e., if a star
has not severely depleted Li at that age, it will not
deplete it afterwards (except again for Li--poor stars in M~67) until first
dredge-up and dilution occur ; {\it ii)} the clusters have all the same initial
abundance, but different initial conditions and/or different parameters
led to different Li depletion rates, with NGC~188 being
characterized by the longest Li depletion timescale.
Since on theoretical grounds metallicity, and more in general $\alpha$-element
abundances, are thought to affect both standard
and non-standard mixing processes (e.g., Chaboyer et al. \cite{cha95};
Swenson \& Faulkner \cite{faulk}; Piau \& Turck-Chi\`eze
\cite{pt02}), a difference in chemical composition could be
a possible cause for different timescales of Li depletion; {\it iii)}
the clusters have different initial Li abundances; in particular, the very old
cluster NGC~188 might have had, because of its age, a lower Li abundance than 
the younger clusters. In this case, it must have also undergone considerably
less Li depletion than the other clusters to end up with the same average
Li abundance.

At present we have no definite clues to discern between these three
possibilities; hypothesis {\it i)} is the simplest one, but requires
a physical mechanism that is not efficient at ages older than about 2 Gyr.
None of the Li depletion processes so far proposed has these characteristics.
As to the second scenario, we note that no Li depletion -- metallicity
relation is found for solar-type stars up to the Hyades age (e.g., Jeffries
\cite {jef00} and references therein). Furthermore,
Pasquini et al. (\cite{pas01})
found  [Fe/H]$=-0.17 \pm 0.11$ for NGC~3680, while Bragaglia et al.
(\cite{brag01}) measured [Fe/H]$=+0.16$ for IC~4651; the two clusters hence
differ in metallicity by a factor of 2, yet they have the same average 
Li abundance. This suggests 
that overall metallicity has little effects on Li depletion also at old ages,
at least when the rather narrow range of [Fe/H] values covered
by Pop.~{\sc i} stars is considered.
In any case, our analysis confirms
that NGC~188 has  solar metallicity and solar metallicity
has also been reported for M~67 (e.g., Jones et al. \cite{jon99} and discussion
therein); thus, even if metallicity would affect Li depletion,
it is not plausible that the two clusters had
different Li histories due to different overall metallicities.
Besides iron, the abundance of $\alpha$ elements
and in particular of oxygen significantly affects stellar opacities,
the depth of convective zone, and in principle mixing (e.g., 
Swenson \& Faulkner \cite{faulk};
Piau \& Turck-Chi\`eze \cite{pt02}). With the exception of
the Hyades, a detailed $\alpha$
element abundance analysis for the clusters shown in Fig.~5 has so far
not been carried out and thus we are not able to investigate
whether the flat \nli~vs. age distribution after 2~Gyr is the result
of different heavy element abundances; given the old
age of NGC~188, it is well possible that the relative abundance
of metals in this cluster may differ from the solar distribution.
However if Li evolution in the four clusters of Fig.~5 older than 1~Gyr
is driven by heavy element abundances,
it would be surprising that these abundances are precisely tuned 
to give the same average Li abundance in all clusters. 
The same argument applies to the third 
possible scenario; whereas
lower initial Li abundance for NGC 188 cannot be excluded (although the
inferred solar metallicity of NGC 188 together with the observed Li vs. Fe
Galactic enrichment argue against this possibility), this
assumption would imply that NGC 188 has also suffered throughout
its lifetime a much lower Li depletion than the younger clusters. We concur
with Hobbs \& Pilachowski (\cite{hp88}) that the possibility that NGC~188
was characterized by a different initial Li
abundance seems very unlikely.

As a final remark, we note that
the average abundance of the 13 NGC~188 members in our
sample is very close to the value of the Li plateau for Pop.~{\sc ii}
halo stars with [Fe/H] $\leq -1.4$
and turn-off stars in globular clusters (e.g., Bonifacio et al. \cite{bon02}).
This point may be a coincidence and Pop.~{\sc ii} stars on
the Spite's plateau cover a wider range of temperatures; thus,
we do not intend
to draw any conclusion from it. However, we regard this coincidence
as very intriguing and worth of further investigation.
\subsection{The dispersion}
Randich et al. (\cite{r00}) from the lack of  dispersion in Li
abundances among solar--type stars of the intermediate age clusters IC~4651 
and NGC~3680,
concluded that the dispersion must have developed after 
$\sim$ 2 Gyr; if this is indeed the case, any cluster older than
that age should exhibit a dispersion. Our results for NGC~188
suggest instead that M~67 might be a peculiar cluster
and that solar--type stars in clusters normally do not develop a dispersion
in Li. A larger number of intermediate-age/old clusters 
is obviously needed, as well as new observations of NGC~752, 
to investigate whether M~67 is really unique and to put this conclusion
on firm basis.
We recall however that a spread in
Li is also present among old stars in the field (e.g., Duncan 
\cite{dun81}; Pallavicini et al. \cite{pal87}; Pasquini et
al. \cite {pas94}): in particular, we mention that several field stars
as old as the Sun, but with much higher Li content exist.
The simultaneous presence of Li-rich and Li-poor stars
in M~67 and in the field implies that, depending
on a parameter that is neither age nor mass, 
Li destruction can be either rather slow or very fast.
Various hypothesis have been proposed in the literature to explain
the star-to-star scatter in M~67; for example, 
the co-existence of two
sub-clusters (e.g. Garc\'\i a L\'opez et al. \cite{gar88}), 
a scatter in initial rotation rates (e.g.,
Jones et al. \cite{jon99}) or, more recently,
a scatter in heavy
element abundances (e.g., Piau et al. \cite{piau03}).

As discussed by Randich et al. (\cite{r00}), if the dispersion observed in M~67
is due to different initial rotation rates and angular
momentum evolutions, the lack of a dispersion in other old clusters
and in particular in NGC~188, would imply that solar--type stars in these
clusters arrived on the ZAMS with very similar initial rotation rates;
this is quite unlikely since
a dispersion in initial rotation rates is indeed observed in all 
the young clusters so far surveyed for rotational
periods and/or velocities (e.g., Stauffer et al. \cite{stau97}; Barnes
\cite{bar00} and references therein).
We also mention that, according to current models including
mixing driven by angular momentum, a scatter in Li abundances at the
age of M~67 would imply a scatter in Be abundances. Randich et al. 
(\cite{R02}) instead measured the same Be abundances for M~67 stars
with a different Li content.

On the other hand,
we do not have observational evidences to
proof or dis-proof the other two possibilities, i.e., whether the
scatter is due to differences in heavy element abundances among M~67
stars or if the cluster results from two different subclusters.
We note however that both hypotheses would imply that the population of
M~67 is not homogeneous, confirming that M~67 is a peculiar cluster.
This would also be in agreement with the fact that a dispersion in
Li is observed among field stars, i.e., within a very 
inhomogeneous sample. 


%
\section{Conclusions}
We have determined Li abundances for 11 G--type stars (with dereddened
B--V colors 0.57 $\leq$ \bmo $\leq 0.65$) in the old open
cluster NGC~188. Our data, together with the Li abundances of five stars
(three in common with us) previously measured by Hobbs \& Pilachowski
(\cite{hp88}), allow us to extend the investigation of Li depletion
in Galactic open clusters to ages 2--4~Gyr older than the Sun.
Two are the main results of this study:\\
{\bf a)} solar--type stars in NGC~188 have all the same Li abundances 
(\nli$=2.34 \pm 0.13$) with virtually no scatter, contrary to what has
been observed in M~67;
{\bf b)} solar--type stars in NGC~188 are no more Li depleted than stars in the 
upper envelope of the Li abundance vs. temperature distribution of
M~67.
We have also estimated the cluster metallicity 
finding [Fe/H]$\simeq 0$. 

The above results represent further
challenges to our interpretation of Li depletion
in MS solar--type stars. Based on the results for M~67
and field stars, one would have expected to find a significant 
fraction of Li-poor stars in NGC~188. Finding that 13 out of 13 members
of NGC~188 have a Li abundance only slightly below the Hyades 
confirms that very old stars are not necessarily Li depleted;
at the same time, the absence of NGC~188 members
as Li depleted as the Sun or as the most depleted stars in M~67,
appears intriguing and raises questions about the possibility
of using a single open cluster as truly representative of all
clusters of the same age and metallicity.

The comparison of NGC~188 with the upper envelope of
M~67, with the 2~Gyr clusters,
and with the Hyades shows that the average Li abundance is virtually
the same for cluster with ages $\geq 2$~Gyr. 
Whereas this result can be due to differences
in heavy element abundances leading to different Li
depletion timescales for different clusters, this possibility appears
to be not very likely. 
We rather believe that the plateau in Li abundances observed for ages
older than 2~Gyr supports the idea that, for part of the stars,
Li depletion may become ineffective beyond this age.
A suitable mixing process with these characteristics
must be investigated. 
The spread among M~67 and in the field shows that part
of the stars instead undergo a very dramatic additional Li destruction
after 2~Gyr. However
M~67 remains so far the only
old cluster for which a dispersion among solar--type stars has
been confirmed. The reasons for this remain unclear,
but we suggest that M~67 might represent an
inhomogeneous sample in some respect.

We finally mention that the average
Li abundance of NGC~188 is only slightly higher than the
Li plateau for Pop.~{\sc ii} stars with [Fe/H] $\leq -1.4$
in the halo and in globular
clusters. Although this may be only a coincidence, it indicates the need 
of determining Li abundances in other old clusters (both open and globular)
in order to link together the Li abundance of Pop.~{\sc ii} stars (usually
assumed to be equal to the primordial value) with that of
Pop.~{\sc i} stars in old open clusters.

\begin{acknowledgements}
We thank the referee, Dr. F. Spite, for very useful comments on the manuscript.
\end{acknowledgements}

{}
\end{document}